%% file: main.tex
\documentclass[10pt,twocolumn,letterpaper]{article}

\usepackage{iccv}              

\usepackage{pgfplots}
\pgfplotsset{compat=1.18}
\usepackage{verbatim}
\usepackage{graphicx}
\usepackage{pifont}

\usepackage{colortbl} 


\definecolor{iccvblue}{rgb}{0.21,0.49,0.74}
\usepackage[pagebackref,breaklinks,colorlinks,allcolors=iccvblue]{hyperref}

\definecolor{gray}{gray}{0.9}
\definecolor{aquamarine}{rgb}{0.5, 1.0, 0.83}
\definecolor{aqua}{rgb}{0.0, 1.0, 1.0}
\definecolor{pink}{rgb}{1.0, 0.75, 0.8}
\definecolor{lightbrightturquoise}{rgb}{0.68, 0.85, 0.9}
\definecolor{brightturquoise}{rgb}{0.03, 0.91, 0.87}

\def\paperID{2231} 
\def\confName{ICCV}
\def\confYear{2025}

\title{AUTV: Creating Underwater Video Datasets with Pixel-wise Annotations}

\author{Quang Trung Truong\textsuperscript{1}
Wong Yuk Kwan\textsuperscript{1}
Duc Thanh Nguyen\textsuperscript{2}
Binh-Son Hua \textsuperscript{3}
Sai-Kit Yeung\textsuperscript{1}\\
\textsuperscript{1}Hong Kong University of Science and Technology
\textsuperscript{2}Deakin University
\textsuperscript{3}Trinity College Dublin}

\begin{document}
\twocolumn[{%
\renewcommand\twocolumn[1][]{#1}%
\maketitle
\begin{center}
    \centering
    \captionsetup{type=figure}
    \includegraphics[width=1\textwidth]{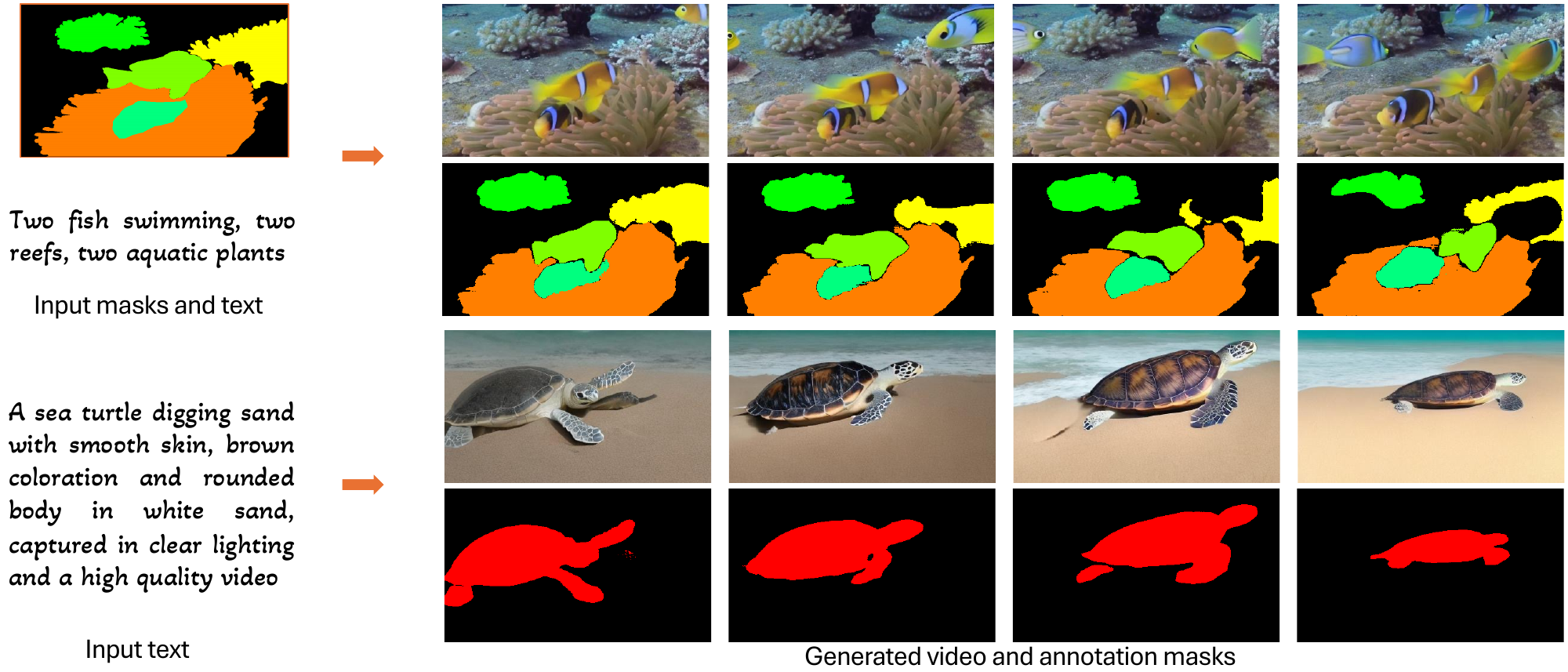}
    \captionof{figure}{Given masks of the first video frame and a text, or a text, our method synthesizes a high-fidelity video}
    \label{fig:teaser}
\end{center}%
}]

\begin{abstract}
Underwater video analysis, hampered by the dynamic marine environment and camera motion, remains a challenging task in computer vision. Existing training-free video generation techniques, learning motion dynamics on the frame-by-frame basis, often produce poor results with noticeable motion interruptions and misaligments. To address these issues, we propose \textsc{AUTV}, a framework for synthesizing marine video data with pixel-wise annotations. We demonstrate the effectiveness of this framework by constructing two video datasets, namely \textsc{UTV}, a real-world dataset comprising 2,000 video-text pairs, and \textsc{SUTV}, a synthetic video dataset including 10,000 videos with segmentation masks for marine objects. UTV provides diverse underwater videos with comprehensive annotations including appearance, texture, camera intrinsics, lighting, and animal behavior. SUTV can be used to improve underwater downstream tasks, which are demonstrated in video inpainting and video object segmentation.

\end{abstract} 

\section{Introduction}
\label{sec:intro}
Approximately 75\% of the Earth's surface (or 362 million $km^2$, equivalently) is dominated by oceans and major seas. These vast bodies of water are integral to climate regulation and serve as a primary source of oxygen for the planet. However, marine species are considerably less well-documented compared with land species. Almost 89\% of marine protected areas are under-explored~\cite{pike2024ocean} and only 200 marine areas are recorded in the World Database on Protected Areas~\cite{roessger2022turning}. 

The underwater computer vision domain aims to interpret marine imagery and video footage. The literature has shown a large body of research methods devoted to underwater video analysis. For instance, underwater instance segmentation is studied in~\cite{lian2023watermask, lian2024diving}. FishNet~\cite{khan2023fishnet} provides a large-scale dataset of fish imagery, supporting various vision tasks such as fish detection and recognition. IOCFormer~\cite{sun2023indiscernible} addresses the underwater object counting problem. MarineInst~\cite{ziqiang2024marineinst} leverages marine data annotation by utilizing both human-annotated and model-generated instance masks for training of instance segmentation models. Depth estimation and underwater image restoration are studied in~\cite{varghese2023self}. 







\begin{figure}[t]
\centering
\includegraphics[width=0.9\linewidth]{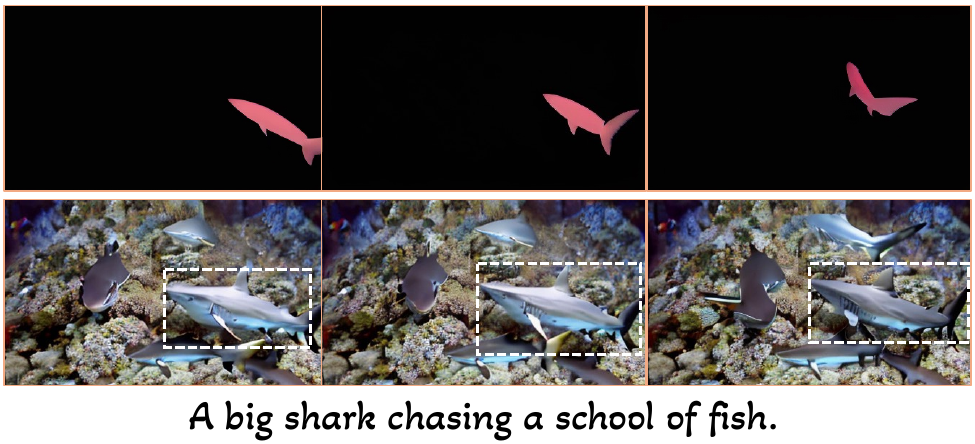}
   \caption{Generated video object masks (first row) and corresponding video frames (second row) by applying SegGen~\cite{ye2023seggen} in a frame-wise basis. As show, there is a huge discrepancy between the generated masks and video frames, due to a lack of integration of motion during the synthesis process.}
   \label{fig:sol1_t2v}
\end{figure}



To facilitate underwater computer vision research, marine datasets have been created. For instance, Marine-tree~\cite{DBLP:conf/cikm/Boone-Sifuentes22a} includes 160k annotated images with 60 object classes organized in a tree structure. UTB180~\cite{alawode2022utb180} contains 180 marine videos for underwater object tracking. MVK~\cite{truong2023marine} consists of 44,330 text-image pairs designed for marine retrieval applications. CBIL~\cite{yifan2024} leverages video-based motion prior to guide a model to learn movement patterns for adversarial imitation. However, CBIL heavily relies on training segmentation masks for motion synthesis. In this paper, we propose a workflow that performs simultaneously video generation and object mask annotation from a textual description or a combination of a textual description and initial masks (see Figure~\ref{fig:teaser}). This capability is enabled by text-to-video (T2V) and object segmentation techniques. 

As shown in the literature, T2V presents a promising approach to creating abundant video content based on textual descriptions. This technology not only improves the richness of marine data, but also supports marine biology studies such as marine species identification and habitat evaluation. A straightforward approach is to leverage text-to-image and mask-to-image methods to video generation on a frame-wise basis. For instance, SegGen~\cite{ye2023seggen} generates object masks for individual video frames using a text-to-image model, fine-tuned on an instance segmentation dataset (e.g., COCO~\cite{lin2014microsoft}). These masks are subsequently used as inputs to a mask-to-image model for video frame generation. However, we observed challenges in this approach. First, it is difficult to ensure temporal consistency in object masks trained with limited data while aligning the masks with generated video frames (see  Figure~\ref{fig:sol1_t2v}). Employing video temporal smoothing techniques could be a treatment to enhance the temporal consistency. However, these techniques alter the generated video content, leading to discrepancies and misalignments between the object masks and the generated video frames. Secondly, video mask generation techniques, trained with video object segmentation (VOS) datasets, such as Ref-YouTube-VOS~\cite{seo2020urvos}, Ref-DAVIS17~\cite{khoreva2019video}, MeViS~\cite{ding2023mevis}, often fail to achieve temporal consistency in the mask generation results as the referring datasets are captured at different frame rates, and hence diverse in motion flow and speed.

To enforce the alignment between the generated video frames and object masks while ensuring the smoothness of the motion flow in the generated contents, we fine-tune an off-the-shelf video diffusion model pre-trained on a large-scale dataset, e.g., WebVid-10M~\cite{bain2021frozen}, to generate the video content and then apply a conditional segmentation model, e.g., SAM2~\cite{ravi2024sam2}, with initial masks in the first frame as the condition, to generate object masks in the following frames. In our method, we specifically focus on effectively incorporating motion dynamics into the image-to-video synthesis. In summary, we make the following contributions.



\begin{itemize}
    \item We propose \textsc{AUTV}, a T2V framework for synthesizing marine videos and pixel-wise annotations (object masks). This framework aims at automated data synthesis at scale.
    


    \item We collect and annotate a real-world video-text pair dataset, namely \textsc{UTV}, for marine research. This dataset is used to fine-tune a generic T2V model for marine video data synthesis.
    

    \item We introduce a synthetic marine video dataset with pixel-wise object annotations using our proposed framework. We name this dataset \textsc{SUTV}.

    \item We demonstrate the usefulness of the \textsc{SUTV} dataset in two downstream applications: video inpainting and video object segmentation.
    

\end{itemize}

\section{Related work}
\label{sec:related_work}

\paragraph{Text-to-Video (T2V)}
Diffusion-based T2V generation models trained on large-scale datasets, e.g., WebVid-10M~\cite{bain2021frozen}, LAION-400M ~\cite{schuhmann2021laion}, have made significant advances in the field. For instance, Make-A-Video~\cite{singer2022make} employed a two-step approach that aligns text-to-image (T2I) and T2V tasks by training a joint foundational model. Following this research direction, Movie Gen~\cite{polyak2024movie} was built on the autoencoder architecture in~\cite{rombach2022high}, and enhanced by incorporating temporal convolutions with conventional spatial convolutions. The advantage of this method is its ability to encode videos of varying lengths. VideoLDM~\cite{blattmann2023align} and ModelScopeT2V~\cite{wang2023modelscope} extended the 2D-UNet architecture in the stable diffusion model~\cite{rombach2022high} to 3D-UNet by including temporal layers. This approach was then adopted in~\cite{TFT2V, wang2023videolcm}. VideoCrafter1~\cite{chen2023videocrafter1} is an open-source high-resolution video generator, built on the latent diffusion model in~\cite{DBLP:journals/corr/abs-2211-13221}, and offering both T2V and image-to-video (I2V) functionalities. Latte~\cite{ma2024latte} applied a series of transformer blocks to the spatio-temporal tokens of training video sequences to learn the distribution of the video data in a latent space.

\begin{figure*}[ht]
    \centering
    \begin{subfigure}[b]{0.67\textwidth}
        \centering
        \includegraphics[width=\textwidth]{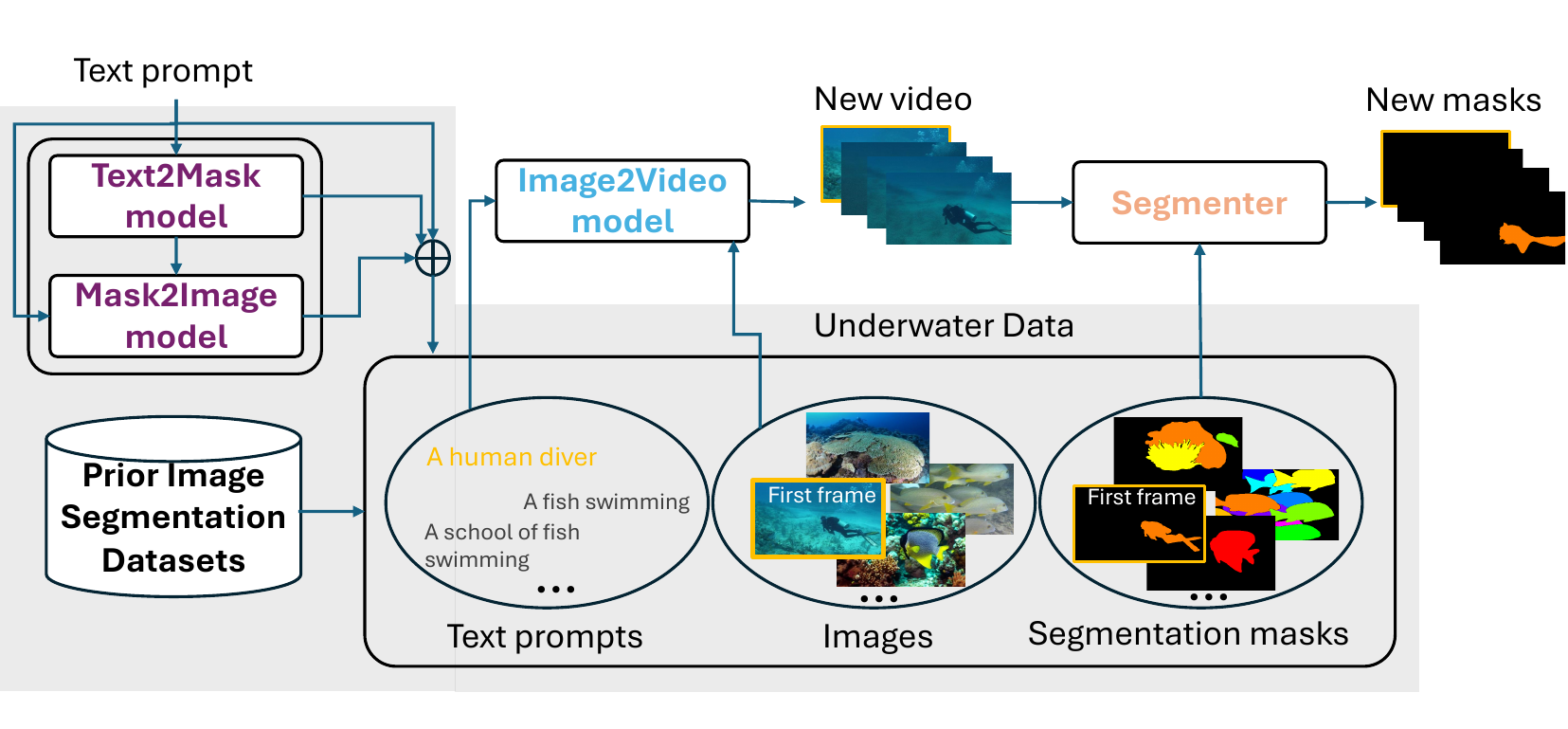}
        \caption{}
        \label{fig:subfig_a}
    \end{subfigure}
    \hfill
    \begin{subfigure}[b]{0.3\textwidth}
        \centering
        \includegraphics[width=\textwidth]{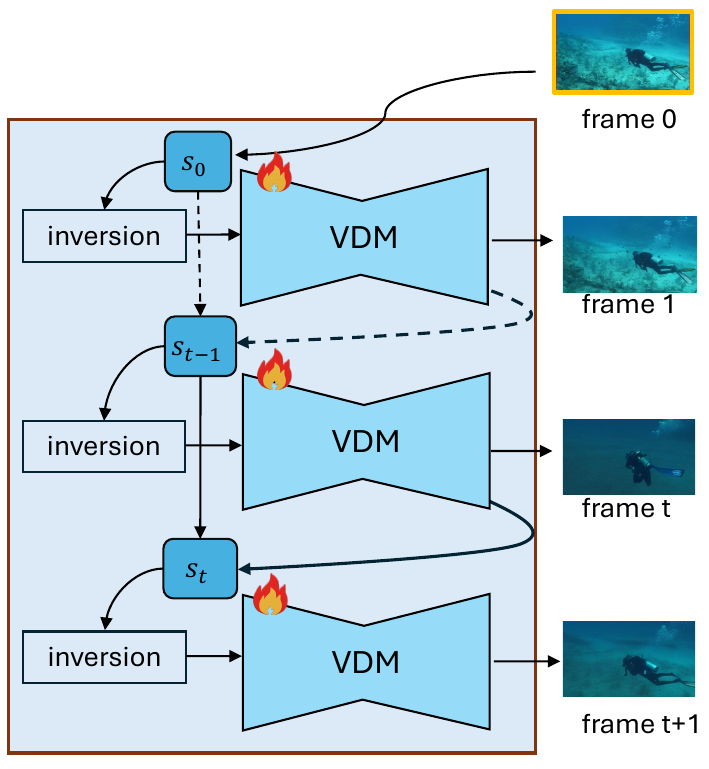}
        \caption{}
        \label{fig:subfig_b}
    \end{subfigure}
    \caption{Overview of our AUTV framework. (a) Video generation pipeline. (b) Image2Video module built on a video diffusion model (VDM), i.e.,  ModelScopeT2V~\cite{wang2023modelscope}. Please zoom-in for the best view.}
    \label{fig:architecture_overview}
\end{figure*}

\paragraph{Training-free T2V}
Training-free methods that extend an image foundational model to video generation provide an efficient solution. This success is achieved by leveraging the generalization ability of T2I diffusion models trained on larger-scale imagery data to smaller-scale video counterparts, producing high-fidelity videos with coherent video-text alignment. For instance, ControlVideo~\cite{DBLP:conf/iclr/Zhang0J0Z024} extended ControlNet~\cite{zhang2023adding} by adding cross-frame interaction to the self-attention module and an interleaved frame smoother to stabilize the denoising process. ConditionVideo~\cite{DBLP:conf/aaai/PengC0LQ24} added a 3D branch to learn the temporal representation of a condition (e.g., a foreground object). BIVDiff~\cite{shi2024bivdiff} demonstrated that the training-free approach can excel in several downstream tasks, including controllable video generation, video editing, and video inpainting and outpainting. However, we observed that this method often struggles to control motion information in the marine domain, probably because of the dynamic of the marine environment. 

\section{AUTV framework}
\label{sec:methodology}

\subsection{Overview}

We introduce \textsc{AUTV}, a framework for synthesizing videos and pixel-wise annotations (object masks) from text prompt conditioning. We build our framework on state-of-the-art diffusion-based T2V and object segmentation techniques. Specifically, AUTV takes input as a text prompt and passes it to a text-to-mask (T2M) model to generate object masks of the first frame in a video sequence. These object masks and input text prompt are subsequently forwarded to a mask-to-image (M2I) model to synthesize the first image frame. The first image frame is then inputted into an I2V model considering both spatial and temporal consistency to generate an output video. This output video and the object masks in the first frame are fed to a segmentor to result in a sequence of object masks. We depict our framework in Figure~\ref{fig:architecture_overview} and describe the main components of the framework in detail in the respective subsections below.

\subsection{T2M generation}
Input of the T2M generation is a text prompt and the output is the object maps of marine objects in the first video frame. We customize and fine-tune Stable Diffusion-v1.5 (SD)~\cite{rombach2022high} in the marine domain to generate object masks of marine species. SD has shown its effectiveness in generating high-quality images from textual inputs and is considered the state-of-the-art in T2I generation. Our rationale for choosing SD is to generate pixel-wise annotations that play a role as a condition to guide the subsequent video frame generation process. We first fine-tune SD for the task of T2M generation on referring expression video object segmentation benchmark datasets including Ref-YouTube-VOS~\cite{seo2020urvos} and Ref-DAVIS17~\cite{khoreva2019video}. We then fine-tune SD to adapt it to the marine domain using our real-world marine dataset, including video-text pairs (presented in Section~\ref{sec:UTV}).


\subsection{M2I generation}

Conditional image generation aims at synthesizing images based on user-provided signals, e.g., object masks. Here we generate an image whose content aligns with a given set of object masks and a text prompt. We apply ControlNet~\cite{zhang2023adding}, built on SD~\cite{rombach2022high}, for conditional image generation, where input is a text prompt and condition is a set of object maps. The output of the M2I generation is an image whose semantic content and generated objects are specified in the input text prompt and align with the input object maps, respectively. This image is considered the first frame of an output video sequence.


\subsection{I2V generation} 
\label{sec:t2v_model}

The I2V generation receives the first video frame and generates the remaining frames of a video sequence. Initially, we adopt a pre-trained diffusion-based T2V model, e.g., ModelScopeT2V~\cite{wang2023modelscope}, trained on large-scale datasets as our baseline. To adapt the model with the marine domain, we fine-tune the model using our real-world dataset (presented in Section~\ref{sec:UTV}). 

Diffusion-based T2V can be performed in a frame-wise approach. For instance, BIVDiff~\cite{shi2024bivdiff} applies Denoising Diffusion Implicit Models (DDIM) inversion~\cite{DBLP:conf/iclr/SongME21} to adapt an image diffusion model for video diffusion. However, synthesizing an entire video directly using a T2V model often fails to enforce temporal consistency. Here, we incorporate the first generated frame as a video prior into frame-wise video diffusion results. Specifically, we keep the first frame and its object masks unchanged. Thus, we implement a frame-by-frame generation approach during each diffusion model sampling process. 

Formally, we represent a video including $M+1$ frames at the $t$-th timestep as $s_t=[x_0,.., x_M]$, where $x_0$ is synthesized using text-guided, mask-conditioned generation with ControlNet~\cite{zhang2023adding} (i.e., the T2M and M2I modules). This frame is duplicated for the remaining $x_i, i>0$ to achieve a synthesized video of length $M + 1$. We use an encoder $\mathcal{E}$ to convert the frame-wise data $s_0$ from the pixel space to the latent space as $z_0$. In the forward diffusion process, a Markov chain $z_1,.., z_T$ is produced by iteratively adding Gaussian noise to $z_0$. The reverse denoising process utilizes a UNet of the fine-tuned ModelScopeT2V~\cite{wang2023modelscope} to gradually reduce noise in the Markov chain $z_{T-1}$,..,$z_0$. We apply DDIM inversion~\cite{DBLP:conf/iclr/SongME21} to initialize the latent representation of $x_i$ from $x_{i-1}$. Here, we hypothesize that frames $x_i$ and $x_{i-1}$ are closely aligned to reinforce the temporal consistency of the generated video. Finally, a decoder $\mathcal{D}$ is used to decode the denoised latent representation of $x_{M+1}$. We illustrate this I2V module in Figure~\ref{fig:architecture_overview}(b). 

\subsection{Video mask generation}
\label{sec:video_annotation_generator}

The video mask generation aims to generate object masks for an entire video using the object masks given in the first frame. SAM2~\cite{ravi2024sam2} requires users to explicitly provide pixel-wise annotations as prompts to guide segmentation. Grounding DINO~\cite{ren2024grounding} uses bounding boxes as prompts for SAM2. However, we found that objects generated by DINO can result in inaccuracies such as false positives or incomplete segmentation (see Figure~\ref{fig:qualitative_ours_baseline}). Here, we use masks generated by the T2M module as prompts for SAM2. We found that this approach works well in marine video segmentation, enabling a better alignment between the generated video contents and the object masks.

\begin{table*}
    \centering
    \scalebox{0.9}{
        \begin{tabular}{l |c|c|c|c|c|c|c|c}
            \toprule
            Dataset & Time & Domain & \#Sentence & Caption & \#Video-text & Cap Source & Attributes & Prompt \\
            & (hours) &&&Length&pairs&&&Complexity \\
            \midrule
            InternVid-VTT~\cite{wang2023internvid} & 76K & open & 1 & 10-20 & 7.1M & synthesis & \ding{55} & medium \\
            MSR-VTT~\cite{xu2016msr} & 41.2 & open & 1 & 9.3 & 200K & human & \ding{55} & low \\
            WebVid~\cite{bain2021frozen} & 13K & open & - & 12 & 2.5M & alt-texts & \ding{55} & low \\
            EPIC-KITCHENS-100~\cite{damen2022rescaling} & 100 & cooking & 1 & 3 & 19.8K & human & \ding{55} & low \\
            \midrule
            \textsc{UTV} (Ours) & 18.48 & \textbf{underwater} & \textbf{3.4} & \textbf{44} & 2K & human & \ding{51} & \textbf{high} \\
            \bottomrule
        \end{tabular}
    }
    \caption{Comparison with existing text-video datasets. }
    \label{table:description_datasets}
\end{table*}

\section{Datasets}
\label{sec:collecting_annotations}


\subsection{UTV - a real-world video-text dataset with fine-grained annotations}
\label{sec:UTV}

Existing datasets such as WebVid~\cite{bain2021frozen} contain watermarked videos and require licenses to fully distribute videos to the community, and underwater videos from YouTube often consist of multiple segments within a single long video. Here, we introduce a real-world dataset, including 2,000 video-text pairs in the marine domain.  

\begin{figure*}[ht]
\begin{center}
\centering
\includegraphics[width=0.85\textwidth]{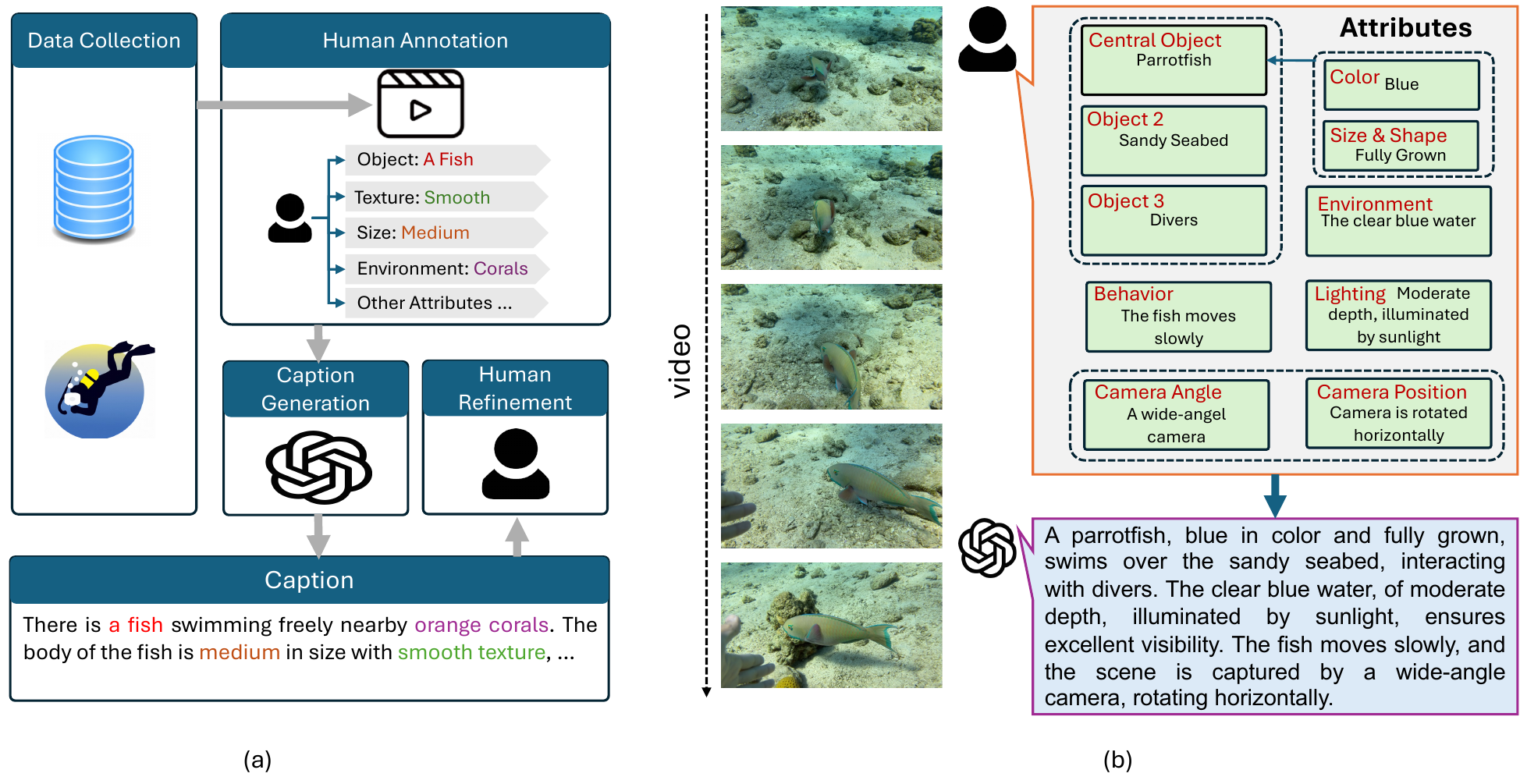}
\end{center}
   \caption{Our video captioning annotation. (a) Our captioning pipeline. (b) An example of annotated attributes. Please see the supplementary material for more examples.}
   \label{fig:datapipeline}
\end{figure*}

\begin{figure*}[htbp]
    \centering
    \begin{minipage}[t]{0.5\textwidth}
        \centering
        \includegraphics[width=1\linewidth]{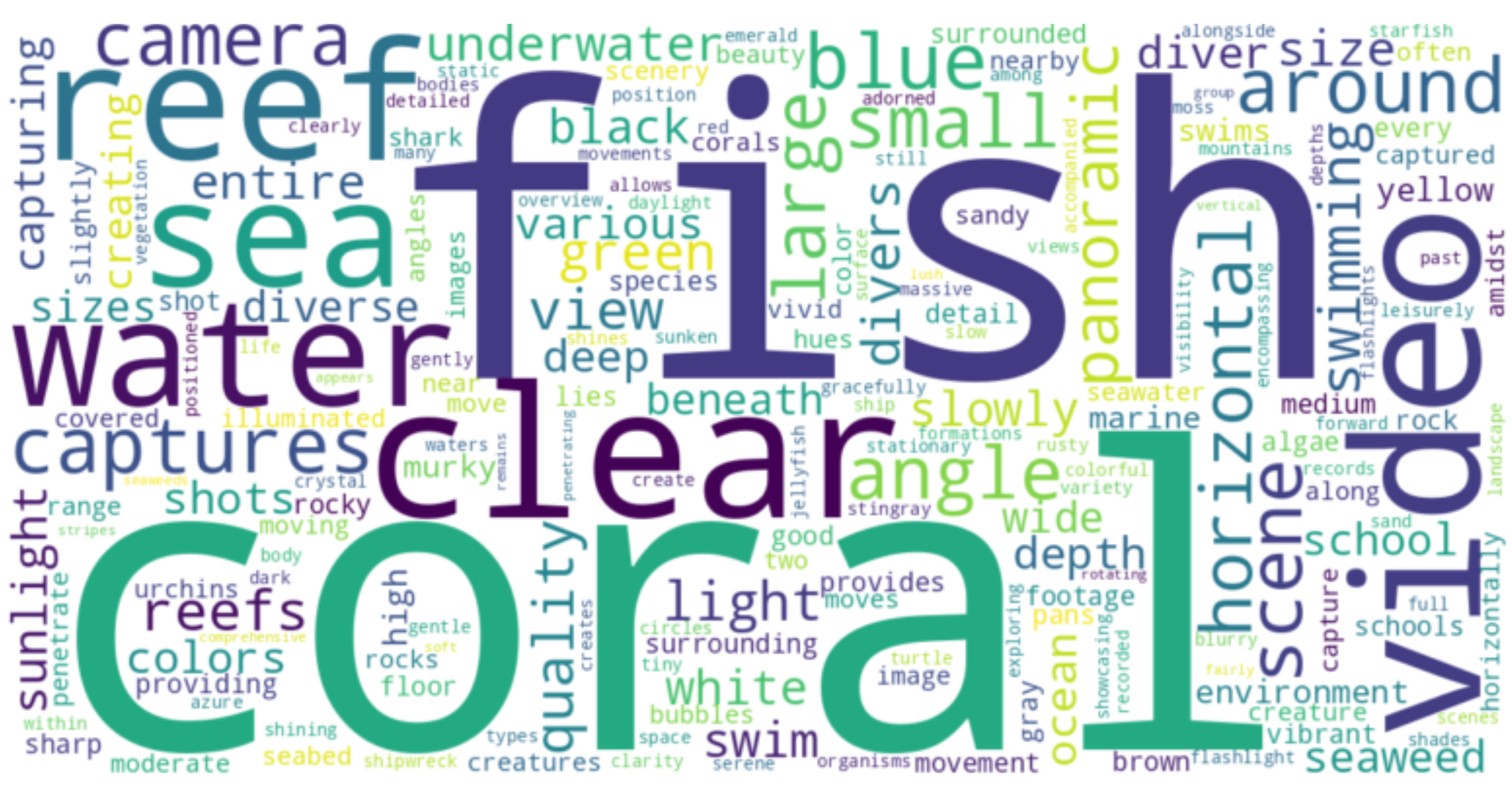}
        \caption{Word cloud of the text in real-world dataset (UTV).}
        \label{fig:caption_wordcloud}
    \end{minipage}%
    \hfill
    \begin{minipage}[t]{0.45\textwidth}
        \centering
        \resizebox{0.96\linewidth}{!}{
            \begin{tikzpicture}
                \begin{axis}[
                    xbar,
                    xlabel={Attribute Percents (\%)},
                    yticklabels={Camera Angle, Camera Position, Movement Behavior, Video Clarity, Lighting, Environment, Object4, Object3, Object2, Size Shape, Color, Texture, Central Object},
                    ytick={1,2,...,13},
                    xmajorgrids,
                    ymax=14,
                    xmax=105
                ]
                \addplot[fill=cyan] coordinates{(100, 1) (100, 2) (82, 3) (100, 4) (50, 5) (100, 6) (24, 7) (52, 8) (88, 9) (96, 10) (91, 11) (11, 12) (100, 13)};
                \end{axis}
            \end{tikzpicture}
        }
        \caption{Statistics on the annotated attributes of our real-world dataset (UTV).}
        \label{fig:attribute_percent}
    \end{minipage}
\end{figure*}

We also provide fine-grained annotations, including \textit{``central object''} and up to \textit{``3 additional objects''}, \textit{``environment''}, \textit{``lighting''}, \textit{``video clarity''}, \textit{``motion/behavior''}, and camera attributes, e.g., \textit{``angle''} and \textit{``position''} for the collected videos. The captioning annotation begins with manually watching each video entirely to identify a central object and any other relevant objects. For each \textit{``central object''}, we further annotate its appearance attributes, including \textit{``texture''}, \textit{``size''}, and \textit{``shape\&color''}. We allow up to 4 objects in each video to be extracted and described in the video's caption. It is important to provide insights of the underwater environment for future ecological studies. To do so, we further describe the environmental factors from the videos including lighting conditions, video clarity, any movement or behavior observed, the camera position (full scene, single angle, partial view of the central object) and video clarity (sharp/blurry), and the camera angle (horizontal, vertical, or centered on the main subject). This captioning process takes around 10 minutes per video, excluding the time required to watch the content. Subsequently, we use ChatGPT to convert the list of attributes into synthetic captions that describe the video content. Finally, a human annotator reviews the generated captions to ensure completeness, remove irrelevant information, and eliminate redundancy or overly generic descriptions. The average caption length is 44 words, covering 13 different attributes. The annotation process and results are illustrated in Figure~\ref{fig:datapipeline}. We compare our UTV with existing video-text paired datasets in Table~\ref{table:description_datasets}. We provide a word cloud for the captions of our UTV and the distributions of the annotated attributes in Figure~\ref{fig:caption_wordcloud} and Figure~\ref{fig:attribute_percent}, respectively.

\begin{table*}
\centering
\scalebox{0.9}{%
    \begin{tabular}{l |ccc|ccc}
    \toprule
        Method & \multicolumn{3}{c|}{FID$\downarrow$} & \multicolumn{3}{c}{FVD$\downarrow$}  \\
        \cline{2-4}\cline{5-7}  
        & $\mathcal{S}$      & $\mathcal{M}$      & $\mathcal{H}$ & FVD64$\downarrow$      & FVD128$\downarrow$      & FVD256$\downarrow$ \\
        \midrule
        Latte ~\cite{ma2024latte}                     & 110.9 & 93.7& 80.4& 4578.1    & 3699.7& 3312.4     \\
        ModelScopeT2V ~\cite{wang2023modelscope}       & 127.4& 101.3 & 92.2  & 3523.3    & 3548.5& 3426.9     \\
        TF-T2V ~\cite{TFT2V}                          & 147.2& 136.8& 128.2 & 4843.5    & 4500.2& 4390     \\  
        VideoLCM ~\cite{wang2023videolcm}             & 167.5& 128.7& 126.2& 3432.5   & 3233.6 & 3018      \\  
        \midrule
        Our \textsc{AUTV}              & \textbf{104.2} & \textbf{87.7} &  \textbf{75.7} & \textbf{1506.9} & \textbf{1401} & \textbf{1303.8}      \\  
    \bottomrule
    \end{tabular}}
\caption{Quantitative evaluation and comparison of our AUTV and existing systems for T2V generation in the marine domain. $\mathcal{S}, \mathcal{M}, \mathcal{H}$ stand for respectively ``Simple'', ``Medium'' and ``Hard''. Best performances are highlighted in bold.}
\label{table:quantitative_res_t2v}
\end{table*}

\begin{figure*}[t]
\centering
\includegraphics[width=1\linewidth]{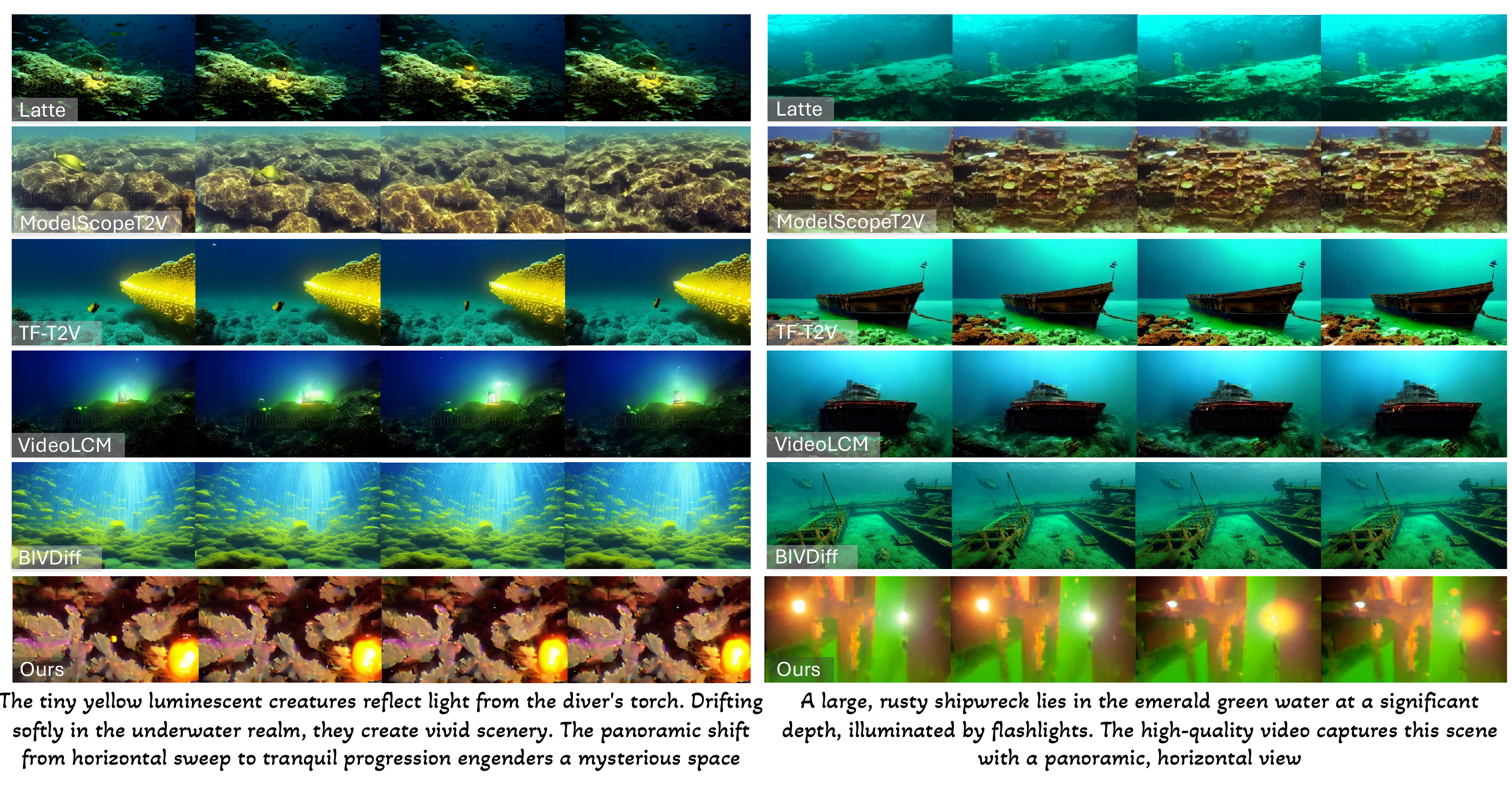}
   \caption{Qualitative evaluation and comparison of our AUTV and existing systems for T2V generation in the marine domain. Our method (last row) demonstrates better text alignment, and more accurately rendering for \textit{``tiny yellow luminescent creatures''} and \textit{``flashlights''}, where the other methods failed. This figure is best viewed in zoomed-in versions.}
   \label{fig:qualitative_res}
\end{figure*}

\subsection{SUTV - a synthetic video dataset with pixel-wise annotations}

We apply our developed AUTV framework to synthesize a marine video dataset with object mask annotations. Specifically, to construct SUTV, we utilize underwater image instance segmentation datasets, including USIS10K~\cite{lian2024diving} and UIIS~\cite{lian2023watermask}, along with their respective categories, to create text prompts. USIS10K comprises 10,632 underwater images with pixel-level annotations, while UIIS contains 4,628 underwater images with annotations. We utilize categories as object attributes defined in our \textsc{UTV} to develop prompt descriptions for generating the synthetic video dataset. We employ ChatGPT to generate prompts, used as input of our AUTV framework.

\begin{figure}
\centering
\includegraphics[width=1\linewidth]{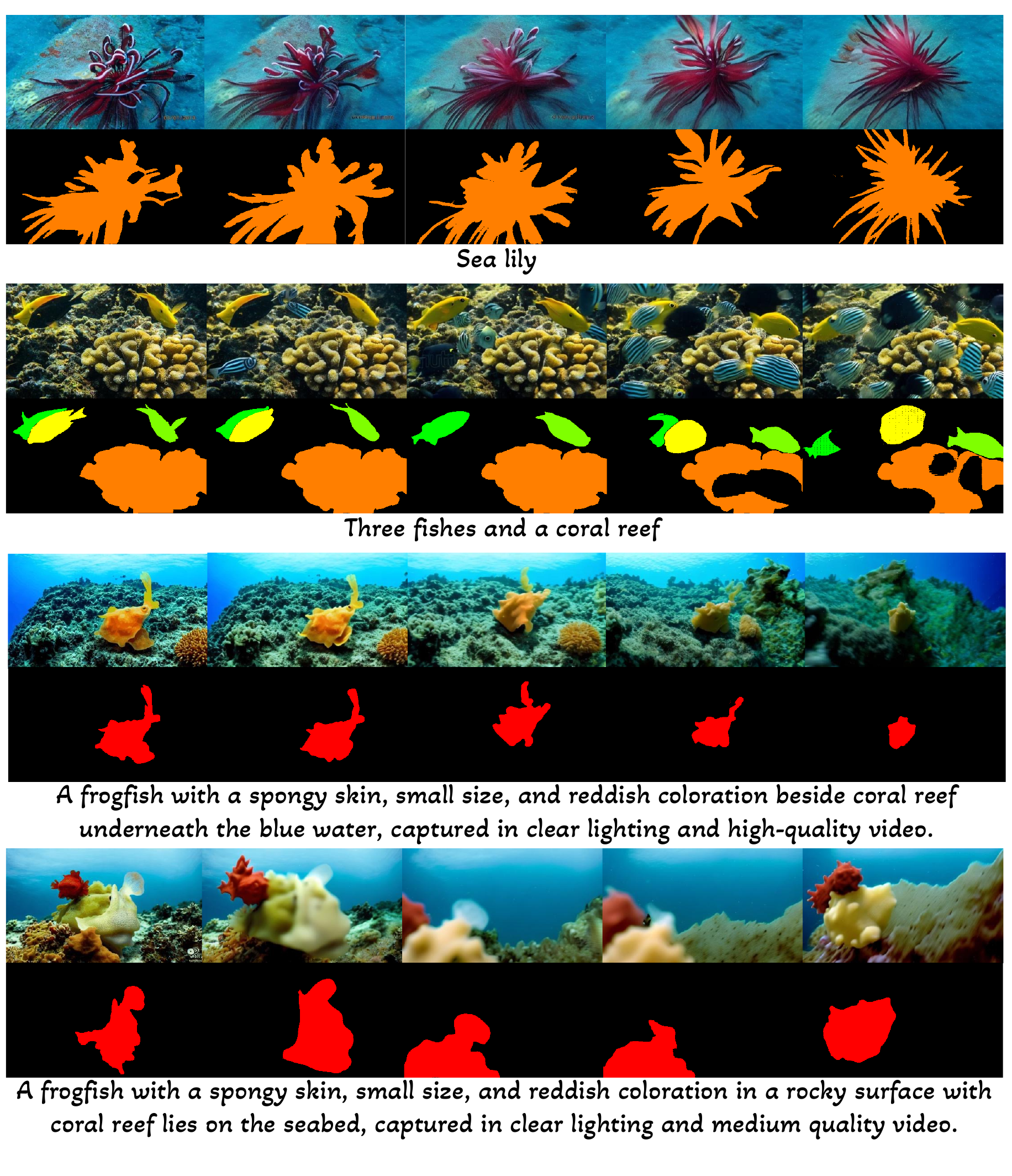}
   \caption{Synthetic videos, segmentation masks and text prompts from our SUTV dataset.}
   \label{fig:dat_examples}
\end{figure}

After generating videos, we apply two filtering steps to the generated videos: motion filtering and visual filtering. \textit{Motion filtering} involves removing videos with frequent jittery camera movements, eliminating those that lack motion, and removing videos featuring special motion effects in both synthesized videos and annotation masks. \textit{Visual filtering} focuses on ensuring no watermarks, minimizing scene changes, and maintaining aesthetic quality in synthesized videos. This process results in a synthetic dataset, including 10,000 video sequences with object segmentation masks. We show several examples of our SUTV in Figure~\ref{fig:dat_examples}.




\section{Experiments}
\label{sec:experiments}


\begin{figure*}
\centering
\includegraphics[width=1\linewidth]{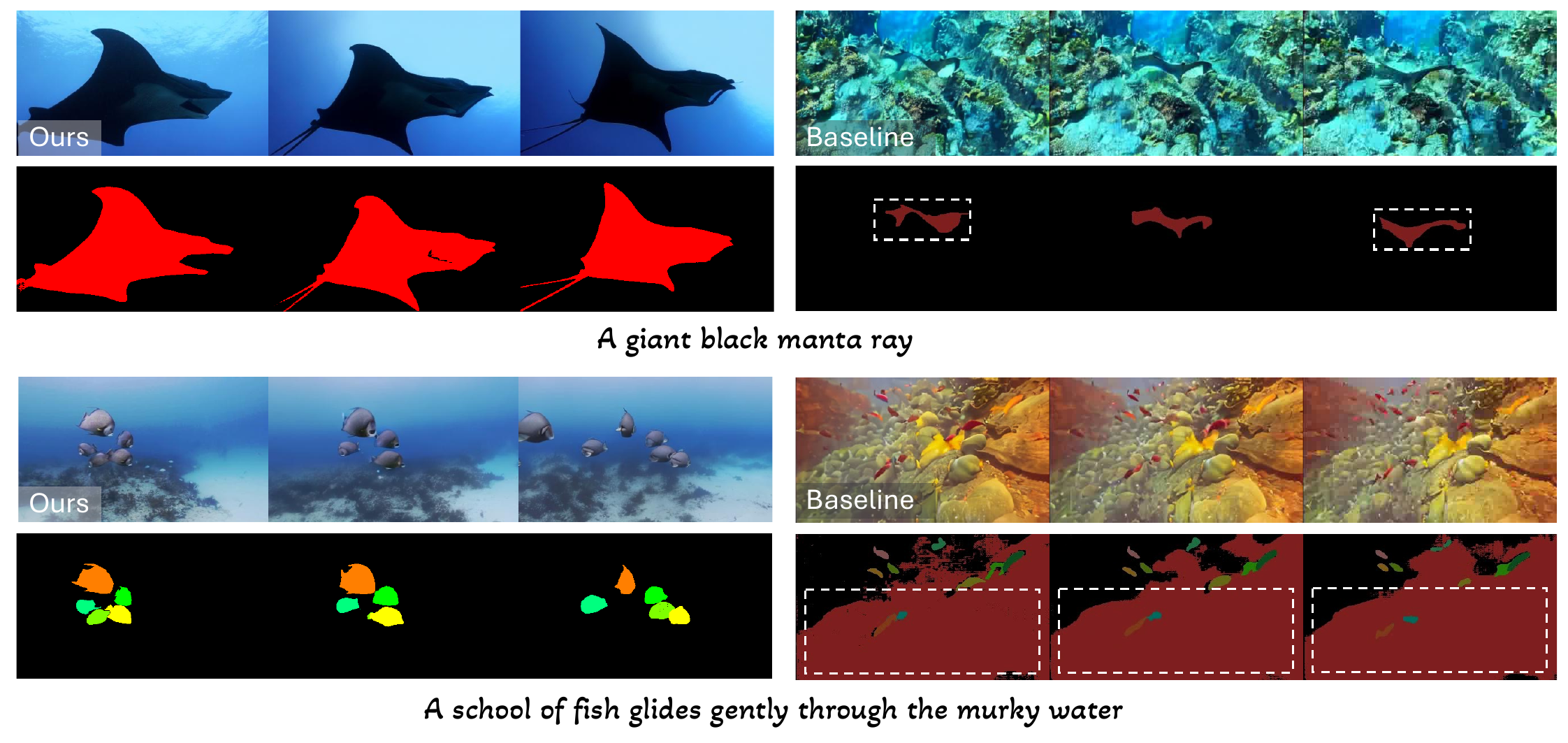}
   \caption{Qualitative mask generation results of our method and a baseline built on ModelScopeT2V~\cite{wang2023modelscope} for I2V and Grounded-SAM-2~\cite{ren2024grounding} for object mask generation. As shown, the baseline fails to produce high-fidelity video frames, resulting in \textit{partial segmentation} (in the first and second rows). Additionally, DINO's limitations in the marine domain lead to inaccurate box prompts, resulting in \textit{false positives} predicted by SAM2~\cite{ravi2024sam2} (in the third and fourth rows). Our AUTV demonstrates superiority in producing high-fidelity video results and annotations for the first frames (as SAM2 prompts), thereby enhancing the alignment between video and annotation masks.}
   \label{fig:qualitative_ours_baseline}
\end{figure*}

\subsection{Experimental setup}
\label{label:implementation_details}
We conducted our experiments, including models fine-tuning, video generation on NVIDIA L20 GPUs. We fine-tuned ModelScopeT2V~\cite{wang2023modelscope} for our I2V generation and SD~\cite{rombach2022high} for our T2M generation in 6 days and 9 days, respectively. To fine-tune ModelScopeT2V for the marine domain, we utilized our collected and annotated real-world dataset (\textsc{UTV}). We categorize \textsc{UTV} based on the complexity of the videos into three levels of difficulty: simple, medium, and hard. Simple videos feature a single central object, medium videos contain two objects within the scene, and hard videos include more than two objects. Our \textsc{UTV} comprises 177 simple videos, 742 medium videos, and 1081 hard videos. To fine-tune SD, we used referring video object segmentation datasets including Ref-DAVIS17~~\cite{khoreva2019video} with 60 videos, Ref-YouTube-VOS 2018 and 2019~~\cite{seo2020urvos} with 2972 videos, and our \textsc{UTV} with 92 videos.

\begin{table}[!t]
\centering
\scalebox{0.9}{
\begin{tabular}{c|cc}
\toprule
Segmentation Metric & Mean & Std. dev. \\
\midrule
mIOU & 89.72 & 6.04 \\
\bottomrule
\end{tabular}}
\caption{Segmentation accuracy of our segmenter.} 
\label{table:segmenter_evaluation}
\end{table}

\subsection{Evaluation of T2V synthesis}

We evaluated our proposed AUTV framework for T2V synthesis both quantitatively and qualitatively. For quantitative evaluations, the Fréchet Video Distance (FVD) ~~\cite{unterthiner2018towards} and the Fréchet Inception Distance (FID) ~~\cite{heusel2017gans} were used as performance metrics for the T2V synthesis. We adopted the evaluation protocol used in StyleGAN-V~\cite{skorokhodov2022stylegan}. The evaluation protocol involves an initial step of sampling the video data and randomly selecting fixed-length video clips from the real data to compute the metrics. We also compared our method with existing T2V baselines, including ModelScopeT2V~\cite{wang2023modelscope}, Latte~\cite{ma2024latte}, TF-T2V~\cite{wang2024recipe}, VideoLCMVideoLCM~\cite{wang2023videolcm}, and BIVDiff~\cite{shi2024bivdiff}.

We report the quantitative and qualitative results of our AUTV and other T2V baselines in Table~\ref{table:quantitative_res_t2v} and Figure~\ref{fig:qualitative_res}, respectively. We observed that existing T2V baselines demonstrate strong performance in open-domain benchmark datasets, e.g., MSR-VTT~\cite{xu2016msr}, SkyTimelapse~\cite{xiong2018learning}, encounter challenges in aligning with text descriptions in the marine domain. The challenges increase accordingly with the number of target objects in an input text prompt.

\subsection{Evaluation of object mask generation}

To verify the annotation quality of our segmenter in object mask generation, we compared object masks generated by our method with manually labeled annotations. Specifically, we randomly selected 100 videos from our synthetic dataset (SUTV) and sampled two frames (the 3rd and 7th frames) per video. We used mIOU as the performance metric for the object mask generation. We report the performance of our segmenter in Table~\ref{table:segmenter_evaluation}. As shown in the results, the annotation quality generated by our segmenter achieves 89.72\% mIOU, highlighting the high-quality segmentation masks predicted by our framework. We compare our method with a baseline built on ModelScopeT2V~\cite{wang2023modelscope} for I2V and Grounded-SAM-2~\cite{ren2024grounding} for object mask generation in Figure~\ref{fig:qualitative_ours_baseline}.



\begin{table*}[!t]
    \centering
    \scalebox{0.9}{
        \begin{tabular}{l|c|ccc|ccc|ccc}
            \toprule
            Setting & Iteration & \multicolumn{3}{c|}{YouTubeVOS} & \multicolumn{3}{c|}{DAVIS 2016} & \multicolumn{3}{c}{DAVIS 2017} \\ 
            \cline{3-5}\cline{6-8}\cline{9-11} 
            & & PSNR $\uparrow$ & SSIM $\uparrow$ & VFID $\downarrow$ & PSNR $\uparrow$ & SSIM $\uparrow$ & VFID $\downarrow$ & PSNR $\uparrow$ & SSIM $\uparrow$ & VFID $\downarrow$ \\ 
            \midrule
            \cite{zhou2023propainter} + real data & 700K & \textbf{33.86} & \textbf{0.9713} & \textbf{0.084} & 22.90 & 0.8389 & \textbf{0.946} & 22.00 & 0.7965 & 1.141 \\ 
            \midrule
            \cite{zhou2023propainter} + SUTV & \textbf{60K} & 31.82 & 0.9613 & 0.117 & \textbf{23.26} & \textbf{0.8493} & 1.029 & \textbf{22.24} & \textbf{0.7983} & \textbf{1.111} \\  
            \bottomrule
        \end{tabular}
    }
    \caption{Validation of our SUTV in video inpainting. We train the video inpainter in~\cite{zhou2023propainter} on real-world datasets (DAVIS2016/2017) and our synthetic dataset (SUTV), and compare the respective performances achieved on different training sets. Best performances are highlighted.}
    \label{table:vi}
\end{table*}

\subsection{Downstream applications}

\paragraph{Video inpainting:} 

\begin{figure}[t]
\centering
\includegraphics[width=1\linewidth]{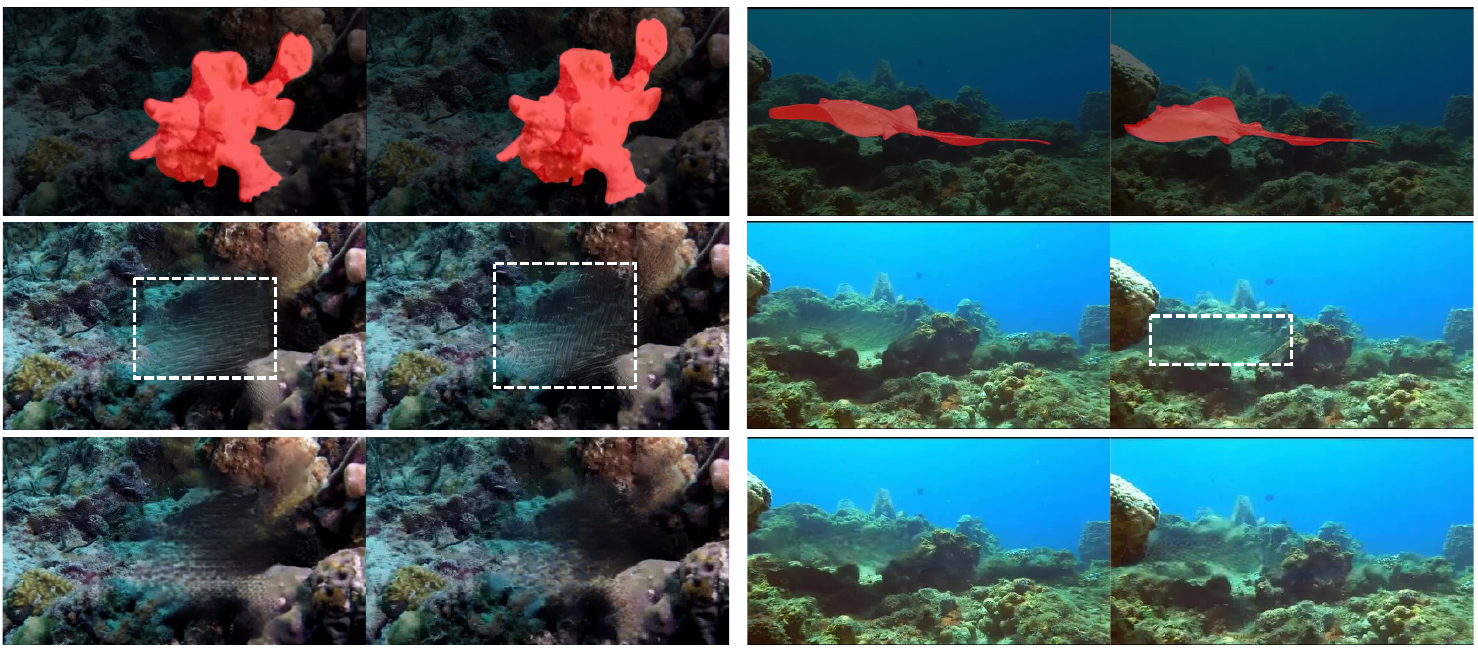}
   \caption{Qualitative results of the ProPainter~\cite{zhou2023propainter} when trained with real-world data (2nd row) and with our synthetic data (3rd row). Masked regions are shown in the 1st row. This figure is best viewed in zoomed-in versions.}
   \label{fig:vi_task}
\end{figure}

We validated our synthetic dataset (SUTV) in the video inpainting application. Specifically, we experimented with the inpainting method in~\cite{zhou2023propainter} in two settings: training with standard datasets (e.g., DAVIS and YTVOS) and training with our SUTV. For a fair comparison, we extracted sub-videos from our SUTV with a fixed length, which is the same as the videos from the DAVIS and YTVOS datasets. Table~\ref{table:vi} demonstrates improved performance achieved by the inpainter in~\cite{zhou2023propainter} when trained with our SUTV over its original version. Additionally, we observed that training on a synthetic dataset leads to significantly faster convergence rates than training from scratch on real-world datasets. Figure \ref{fig:vi_task} illustrates \cite{zhou2023propainter} exhibits superior performance on real-world underwater videos when trained on synthetic data, compared to its counterpart trained on real data.

\paragraph{Video object segmentation (VOS):}
We experimented VOS methods with self-supervised learning paradigm using pseudo labels (object masks) and synthetic masks from our SUTV. To achieve the pseudo labels, we applied the knowledge distillation (KD) techniques in~\cite{truong2024self, huang2022knowledge} to two state-of-the-art VOS frameworks. Specifically, the large variants serve as teachers, while the light models act as students. For the synthetic labels, we used object masks generated by our SUTV. We present results of this experiment in Table~\ref{table:ssl}. We observed that the VOS models trained with our synthetic data significantly outperform their counterparts trained with pseudo labels.

\begin{table}[h]
    \centering
    \scalebox{0.9}{
    \begin{tabular}{l|c|c}
        \toprule
        Method & YouTubeVOS2018 & YouTubeVOS2019 \\
        & $\mathcal{J}$\&$\mathcal{F} \uparrow$ & $\mathcal{J}$\&$\mathcal{F} \uparrow$ \\
        \midrule
        \rowcolor{gray}AOT~\cite{DBLP:conf/nips/YangWY21} & 67.30 & 67.60 \\
        \rowcolor{gray}DeAOT~\cite{yang2022decoupling} & 73.20 & 74.00 \\
        \midrule
        \rowcolor{pink}AOT~\cite{DBLP:conf/nips/YangWY21} & \textbf{74.93} {\small(+7.63)} & \textbf{74.95} {\small(+7.35)} \\
        \rowcolor{pink}DeAOT~\cite{yang2022decoupling} & 74.15 {\small(+0.95)} &     74.14 {\small(+0.14)} \\
        \bottomrule
    \end{tabular}}
    \caption{Quantitative comparisons on self-supervised video object segmentation. We employ a simple KD scheme to distill feature representations and logits from the largest model to the smallest model in gray color. Models trained on the synthetic data are highlighted in red color. We observe that the methods trained on the synthetic data outperform the KD-based models.}
    \label{table:ssl}
\end{table}

\section{Conclusion}
We present \textsc{AUTV}, a framework for synthesizing underwater videos with pixel-wise annotations from text prompts. We collected \textsc{UTV}, a real-world dataset of 2000 video-text pairs from the marine domain. We trained our AUTV using UTV and applied AUTV to generate \textsc{SUTV}, a synthetic video dataset with annotated object masks. Our experiments show that the synthetic dataset can help boost up the performance of various downstream tasks. Our work aims at advancing oceanic research via an automated large-scale data creation method and fine-grained annotated datasets. 



\small
\bibliographystyle{ieeenat_fullname}
\bibliography{main}

\end{document}